\documentclass[a4paper]{article}
\usepackage{graphicx}

\begin{document}

\title{Nuclear astrophysical plasmas: ion distribution functions and fusion rates}
\author{Marcello Lissia \\
Ist. Naz. Fisica Nucleare (I.N.F.N.) Cagliari \\
Dipartimento di Fisica, Universit\`a di Cagliari \\
S.P. Sestu Km~1, I-09042 Monserrato (CA), Italy
\and
Piero Quarati\\
Dipartimento di Fisica, Politecnico di Torino, \\
C.so Duca degli Abruzzi 24, I-10129 Torino, Italy \\
Ist. Naz. Fisica Nucleare (I.N.F.N.) Cagliari}
\date{13 October, 2005}
\maketitle

%
This article illustrates how very small deviations from the
Maxwellian exponential tail, while leaving unchanged bulk
quantities, can yield dramatic effects on fusion reaction rates
and discuss several mechanisms that can cause such deviations.

Fusion reactions are the fundamental energy source of stars and
play important roles in most astrophysical contexts. Since the
beginning of quantum mechanics, basic questions were addressed
such as how nuclear reactions occur in stellar plasmas at
temperatures of few keV (1 keV $\approx 11.6\times 10^6$
${}^\circ$K) against Coulomb barriers of several MeV and what
reactions or reaction networks dominate the energy production. It
was soon realized that detailed answers to such questions involved
not only good measurements or quantum mechanical understanding of
the relevant fusion cross sections, but also the use of
statistical physics for describing energy and momentum
distributions of the ions and their
screening~\cite{Atkinson:1929}.

Gamow understood that reacting nuclei penetrate Coulomb barriers
by means of the quantum tunnel effect and Bethe successfully
proposed the CNO and then the pp cycle as candidates for the
stellar energy production: this description has been directly
confirmed by several terrestrial experiments that have detected
neutrinos produced by pp and CNO reactions in the solar
core~\cite{Castellani:1996cm}.

In the past only few authors (e.g., d'E. Atkinson, Kacharov,
Clayton, Haubold) examined critically the energy distribution and
proposed that such distribution could deviate from the Maxwellian
form. In fact, it is commonly accepted that main-sequence stars
like the Sun have a core, i.e. an electron nuclear plasma, where
the ion velocity distribution is Maxwellian. In the following, we
first discuss why even tiny deviations from the Maxwellian
distribution can have important consequences and then what can
originate such deviations.

\section*{Thermonuclear reaction in plasmas and distribution tails}

In a gas with $n_{1}$ ($n_{2}$) particles of type 1 (2) per cubic
centimeter and relative velocity $v$, the reaction rate $r$ (the
number of reactions per unit volume and unit time) is given by
\begin{equation}
r= (1+\delta_{12})^{-1}\, n_1 n_2 \langle  \sigma v \rangle \ \ ,
\label{rates}
\end{equation}
where $\sigma=\sigma (v)$ is the nuclear cross section of the
reaction. The reaction rate per particle pair is defined as the
thermal average
\begin{equation}
\langle  \sigma v \rangle = \int_0^\infty \! f(v)\,  \sigma v \,
dv \, , \label{sigmav}
\end{equation}
where the particle distribution function $f(v)$ is a local
function of the temperature~\cite{Clayton:1968}.

Therefore, the reaction rate per particle pair $\langle  \sigma v
\rangle$ is determined by the specific cross section and by the
velocity distribution function of the reacting particles. When no
energy barrier is present and far from resonances, cross sections
do not depend strongly on the energy.  Most of the contribution to
$\langle  \sigma v \rangle$ comes from particles with energy of
the order of $k T$, and the dependence on the specific form of
$f(v)$ is weak. The same is true for bulk properties that receive
comparable contributions from all particles: e.g., the equation of
state.

The situation is very different in the presence of a Coulomb
barrier, when the reacting particles are charged, as in the fusion
reactions that power stars~\cite{Coraddu:1998yb}. The penetration
of large Coulomb barriers ($Z\alpha / r$ is of the order of
thousands in units of $k T$ when $r$ is a typical nuclear radius)
is a classically forbidden quantum effect. The penetration
probability is proportional to the Gamow factor
$\exp{[-\sqrt{E_G/E}]}$, where the Gamow energy $E_G= 2\mu c^2
(Z_1 Z_2 \alpha \pi)^2 $, $\alpha$ is the fine structure constant,
$\mu$ is the reduced mass, and $Z_{1,2}$ are the charges of the
ions. The cross section is exponentially small for $E\ll E_G$ and
grows extremely fast with the energy; therefore, one usually
defines the astrophysical $S$ factor, whose energy dependence is
weaker
\begin{equation}
\label{sigmas} \sigma (E)=
           \frac{S(E)}{E} e^{-\sqrt{E_G/E}}  \ \ .
\end{equation}
The two factors in the integrand in Eq.~(\ref{sigmav}) that carry
most of the energy dependence are the Maxwellian distribution
$\propto e^{-(E/kT)}$, which is exponentially suppressed for $E
\gg kT$, and the penetration factor $e^{-\sqrt{E_G/E}}$, which is
exponentially suppressed for $E \ll E_G$. Contributions to the
rate come only from an intermediate region (Gamow peak) around the
temperature-dependent energy
\begin{equation}
\label{mosteffen} E_0 = \left( \frac{E_G (k T)^2}{4} \right)^{1/3}
\, ,
\end{equation}
which is called the most effective energy, since most of the
reacting particles have energies close to $E_0$.

\begin{figure}[hp]
\begin{center}
\includegraphics[width=9cm]{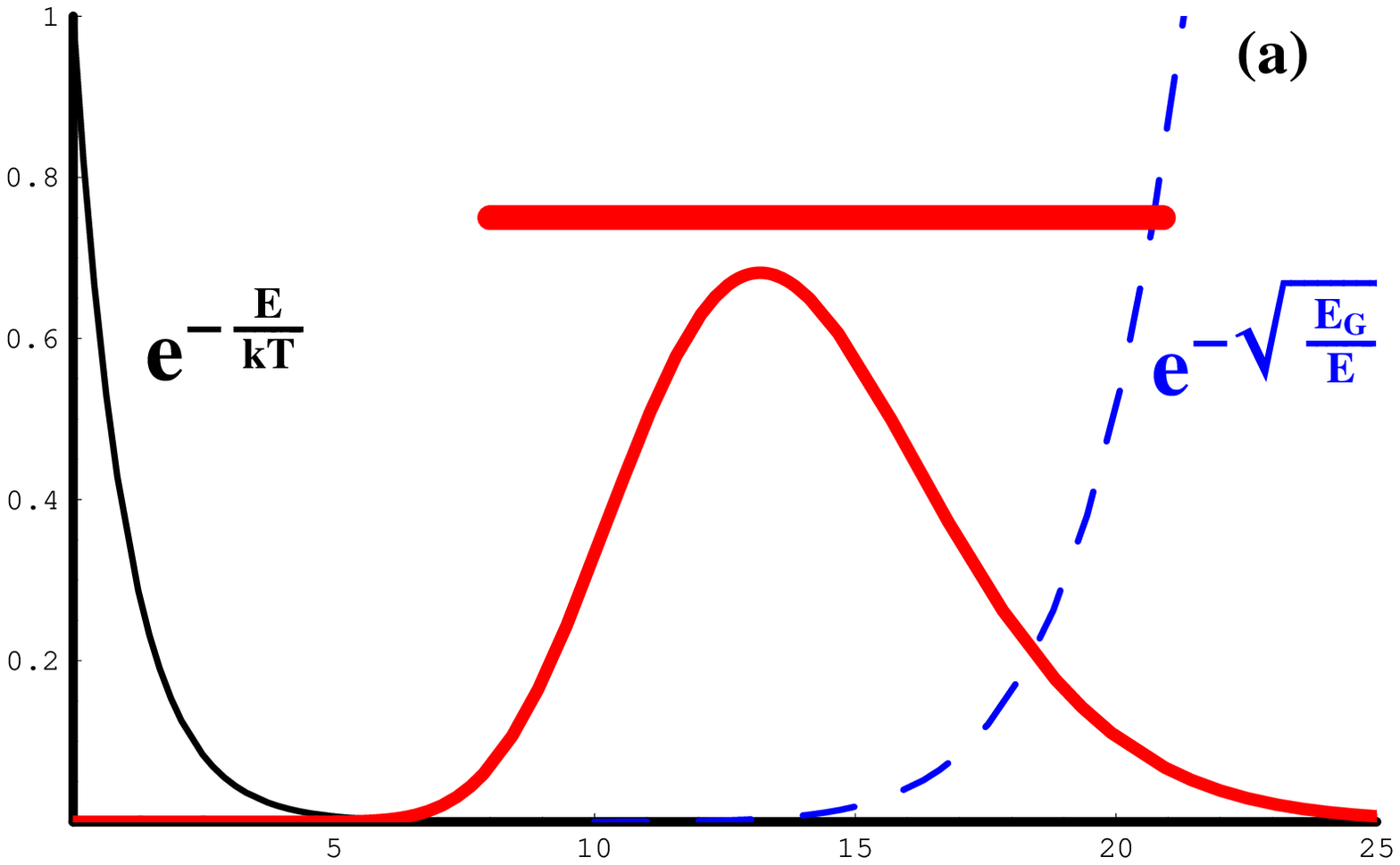}\\
\includegraphics[width=9cm]{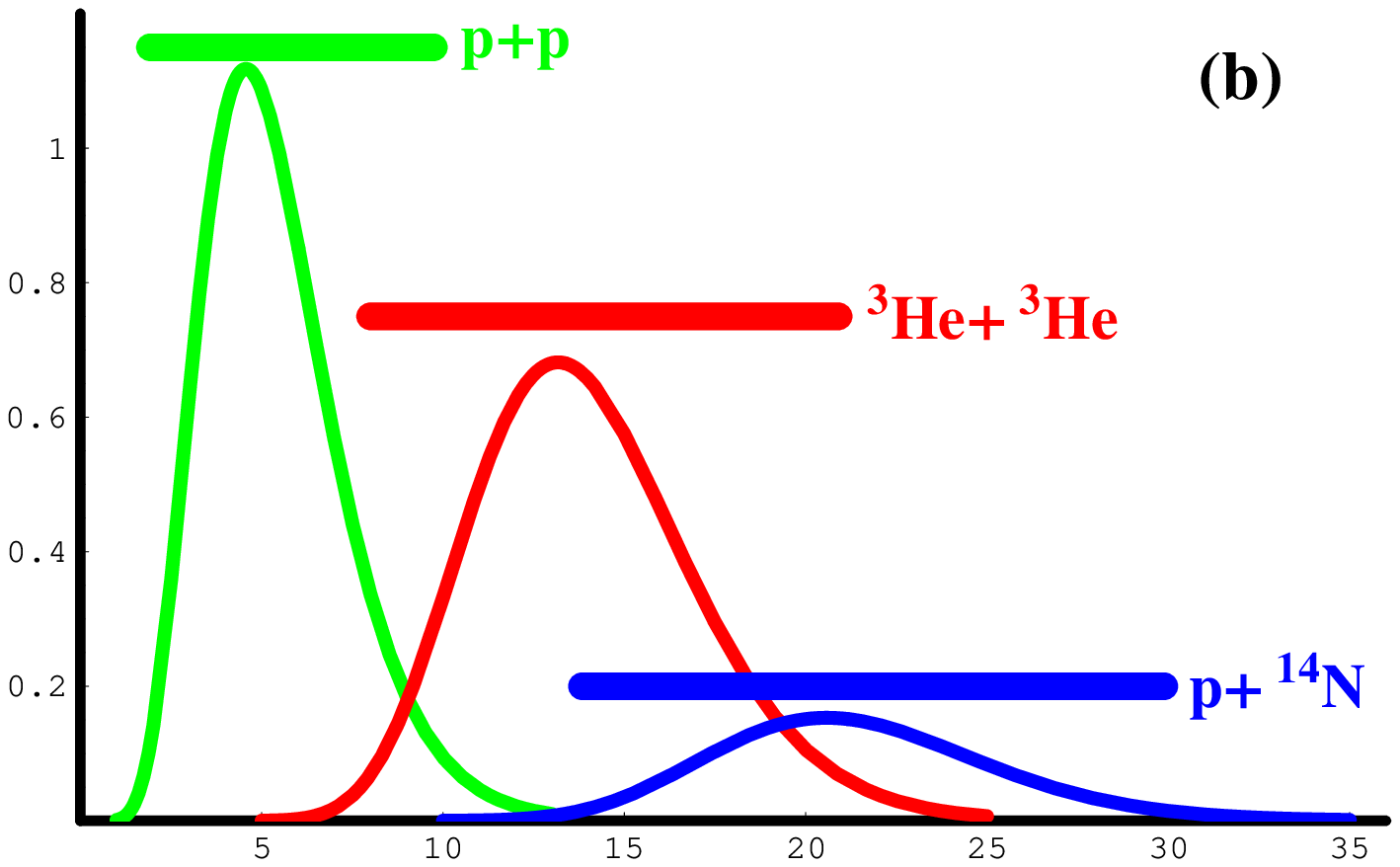}
\caption[fig:gamow]{ \label{fig:gamow} The Gamow peak and energy
selection. In the upper panel (a) the exponential thin black curve
is the Maxwellian distribution. The rapidly increasing dash blue
curve shows the behavior of the penetration factor $N\times
\exp{(-\sqrt{E_G/E})}$ for the solar reaction $^3$He + $^3$He
$\to$ $^4$He + 2$p$ ($E_G= 11.83$~MeV and $N=10^{9}$). The red
thick curve shows the product of the two curves (Gamow peak) times
$10^{8}$. The horizontal red band indicates the energy range of
the reacting particles. The lower panel (b) shows how different
reactions select different windows of particle energies. The Gamow
peak (energy window) moves to higher energies going from  $p+p\to
d + \nu + e^+$ ($E_G= 493$~keV, green), to $^3$He + $^3$He $\to$
$^4$He + 2$p$ ($E_G= 11.83$~MeV, red), and $p + {}^{14}$N $\to$
$^{15}$O + $\gamma$ ($E_G= 45.09$~MeV, blue). Correspondingly, the
peaks become (much) lower; note that the three curves have been
multiplied times $10^{5}$, $10^{17}$, and $10^{26}$, respectively,
to make them visible on the same scale.
  }
\end{center}
\end{figure}

Figure~\ref{fig:gamow} gives a pictorial demonstration of how the
Gamow peak originates and how different reactions select different
parts of the distribution tail and can be used to probe it.

In the upper panel (a) the exponentially decreasing function (thin
black curve) is the Maxwellian factor; the rapidly growing
function (dash blue curve) is the penetration factor (for
graphical reason multiplied by $10^{9}$) of one of the most
important reactions in the Sun, $^3$He + $^3$He $\to$ $^4$He +
2$p$ ($E_G = 11.83$~MeV), which corresponds to a most effective
energy $E_0 = (E_G (kT)^2 / 4)^{1/3} = 17.036$ kT for $kT = 1.293$
keV $ = 11.6\times 10^6$ ${}^\circ$K; the product of the two
functions (Gamow peak) is the thick red curve. Note that the Gamow
peak, and therefore the rate, is very small (it has been
multiplied by an additional $10^{8}$ factor to make it visible on
the same scale of the other curves), since at the most effective
energy $E_0$ both the cross section and the number of particles
are exponentially small. At this point is important to remark that
the area under the Maxwellian curve for energies within the Gamow
peak (the energy window indicated by the red band) is of the order
of 0.1\% of the total area: only a few particles in the tail of
the distributions contribute to the fusion rate.

The fact that the penetration factor effectively selects particles
in the tail of the distribution is the more dramatic the larger
the charge of the reacting ions: for the $p + {}^{14}$N $\to$
$^{15}$O + $\gamma$ (the leading reaction of the CNO cycle, which
dominates the energy production in main-sequence stars larger or
older than the Sun) the contributing particles are few in a
million.

The effect on the Gamow peak when increasing the charges of the
reacting nuclei is shown in the lower panel (b) of
Figure~\ref{fig:gamow}. The green, red, and blue curves show the
Gamow peak  multiplied by $10^{5}$, $10^{17}$, and $10^{26}$,
respectively, for three fundamental reactions in main-sequence
stars: $p+p\to d + \nu + e^+$ ($E_G= 493$~keV), $^3$He + $^3$He
$\to$ $^4$He + 2$p$ ($E_G= 11.83$~MeV), and  $p + {}^{14}$N $\to$
$^{15}$O + $\gamma$ ($E_G= 45.09$~MeV)). It is immediately evident
that the larger the charges of the ions the higher is the energy
of the particles that contribute to the rate,  the (much) lower is
the peak and, therefore, the (much) smaller is the rate. In fact
the maximum of the Gamow peak $E_0\propto E_G^{1/3}\propto
(Z_1Z_2)^{2/3}$ and its hight is proportional to $\exp{(-3
E_0/kT)}$.

A convenient parametrization of deviations from the Maxwell
distribution is the deformed $q$-exponential:
\begin{equation}
\label{eq:expq} \exp_q(-\frac{E}{kT}) = \left(1-(1-q)\frac{E}{kT}
\right)^{1/(1-q)} \, ,
\end{equation}
which naturally appears in Tsallis' formulation of statistical
mechanics~\cite{Tsallis:1987eu}. This particular deformation of
the exponential has the advantage of describing both longer tails
(for $q>1$) and cut-off tails (for $q<1$), while reproducing the
exponential in the limit $q\to1$.

\begin{figure}[hp]
\begin{center}
\includegraphics[width=9cm]{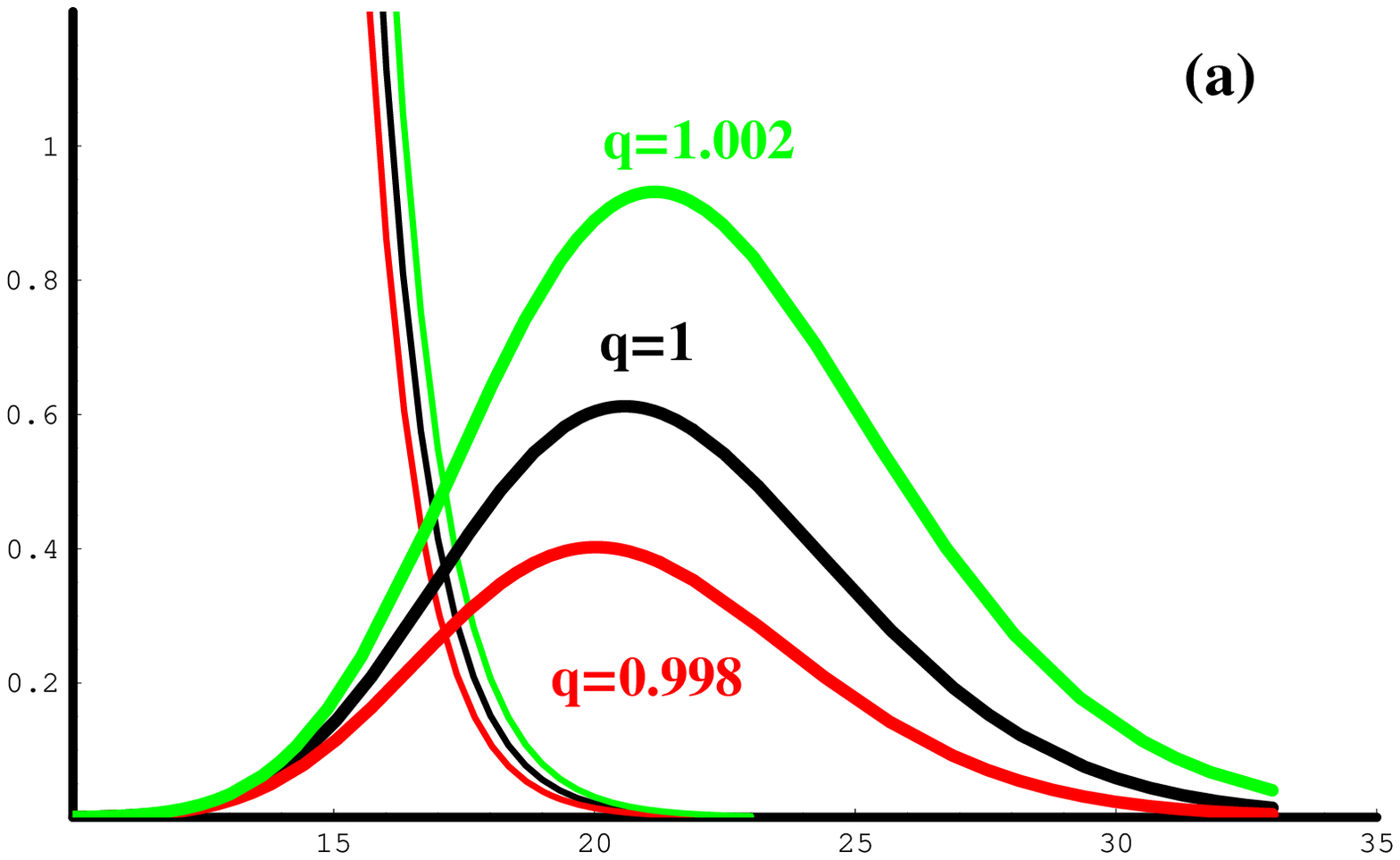}\\
\includegraphics[width=9cm]{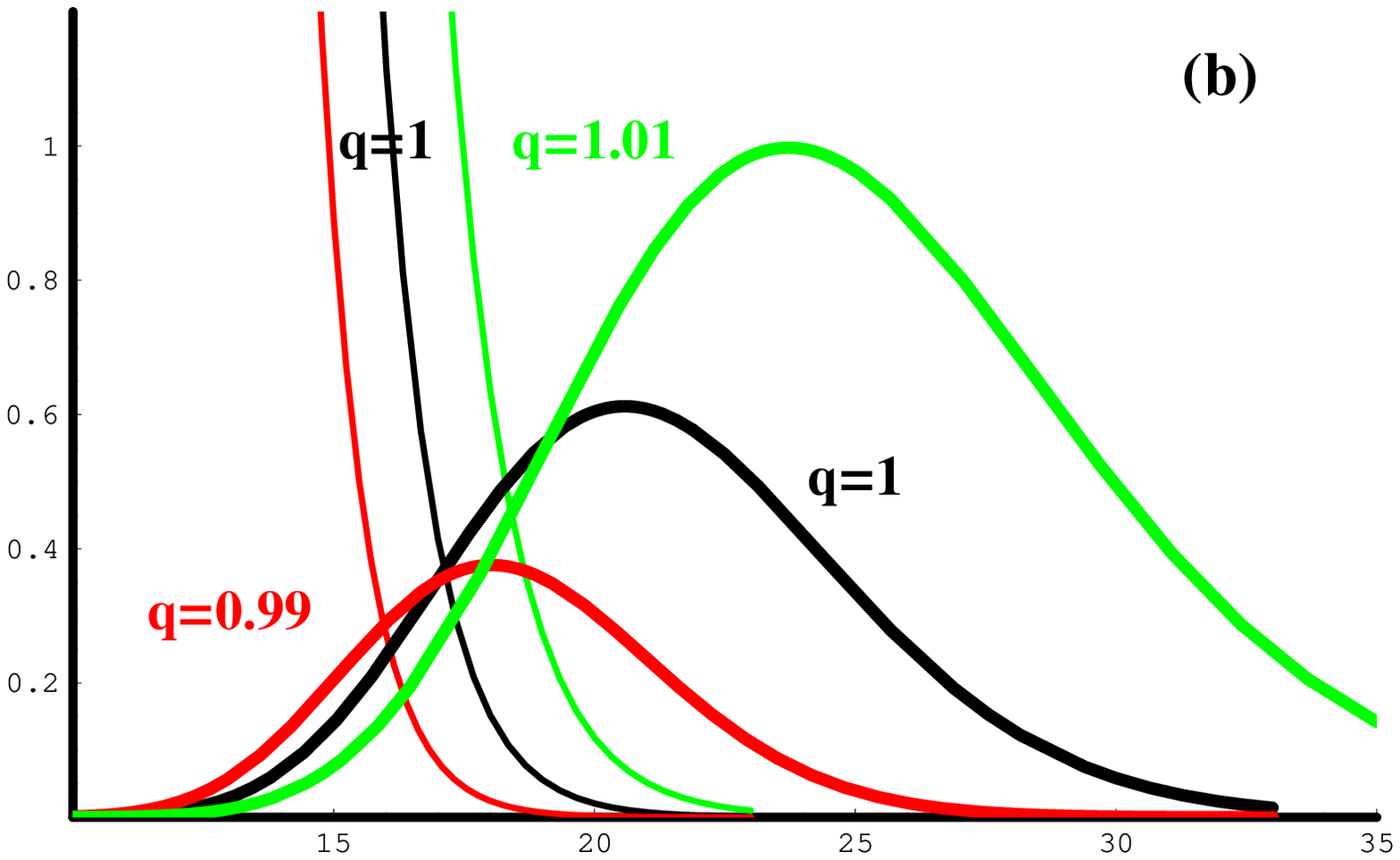}
\caption[fig:defExp]{ \label{fig:defExp} Effects of tiny changes
in the tail of the distribution for the reaction  $p + {}^{14}$N.
Using the parametrization of Eq.~(\ref{eq:expq}), the upper panel
(a) shows the effect of taking $q=1.002$ (green) and $q=0.998$
(red), while the lower panel (b) shows the effect of $q=1.01$
(green) and $q=0.99$ (red); black curves correspond to the usual
Maxwell distribution ($q=1$). In the lower panel (b) the green
(red) Gamow peak (thick lines) is divided (multiplied) by an
additional factor of five. Note that thin curves (exponentials and
q-exponentials) have been multiplied by $10^{9}$ to enounce their
tiny differences.
  }
\end{center}
\end{figure}

Figure~\ref{fig:defExp} shows the effect of substituting to the
Maxwellian distribution $\exp(-E/kT)$ the distribution $ N_q
\exp_q(-E/kT)$, where $N_q$ is the normalization factor that
conserves the total number of particles. We show the effect for
the $p + {}^{14}$N reaction and the values  (a) $q=1\pm 0.002$ and
(b) $q=1\pm 0.01$. This reaction determines the rate of energy
production from the CNO cycle, dominant at older stages and,
therefore, also determines the time of the exit from the main
sequence. Black curves refer to the exponential ($q=1$), red
curves refer to the cut-off ($q<1$) exponential, and green curves
refer to the longer-tail ($q>1$) exponential.

Note that all exponentials have been multiplied times a huge
$10^6$ factor for emphasizing their tiny differences: these values
of $q$ produce functions that are almost indistinguishable from
the exponential unless one looks very far in the tail.

One can make several remarks:
\begin{itemize}
\item
Gamow peaks are shifted towards higher energies, when the
distribution has a tail longer than the exponential (green curves,
$q>1$); they are shifted towards lower energies for cut-off
exponentials (red curves, $q<1$);
\item
the effect is larger the larger the deviation from the exponential
(the larger $|q-1|$);
\item
the peaks (and the rates) become correspondingly  higher for $q>1$
and smaller for $q<1$;
\item
the effect on the rate is already large for $|1-q|=0.002$, it
becomes huge (more than a factor of 10) for $|1-q|=0.01$; note
that the green (red) peak in lower panel (b) of
figure~\ref{fig:defExp} ($|1-q|=0.01$) has been divided
(multiplied) by five to make them fit on the same scale!
\end{itemize}

These deviations should be carefully estimated, since reliable
calculations of nuclear reaction rates in stellar interiors is
fundamental for a quantitative understanding of the structure and
evolution of stars. In fact, while the overall stellar structure
is rather robust, changes of some of the rates even by few percent
can produce detectable discrepancies, when precise measurements
are possible, {\em e.g.}, in the case of the solar photon and
neutrino luminosity, and mechanical
eigenfrequencies~\cite{Castellani:1996cm}. In quasi-stellar
objects like Jupiter deviations could be even larger and explain
their excess energy~\cite{Coraddu:2001ps}.

As already shown in the recent past, very small deviations from
Maxwellian momentum distribution do not modify the properties of
stellar core and are in agreement with the helioseismology
constraints~\cite{Degl'Innocenti:1998dy}, but may affect the
evaluation of the nuclear fusion rates that may be enhanced or
depleted, depending on superdiffusion or subdiffusion property of
the particles~\cite{Coraddu:1998yb}.

\section*{Deviations from Maxwellian distribution}

Normal stellar matter, such as the one in the Sun, is
non-degenerate, {\em i.e.}, quantum effects are small (in fact,
they are small for electrons and completely negligible for ions),
it is non-relativistic, and it is in good thermodynamical
equilibrium. On this ground, the particle velocity distribution is
almost universally taken to be a Maxwell-Boltzmann (MB)
distribution.

Concerning the  thermodynamical equilibrium,  main sequence stars
are more precisely in  a stationary state where the luminosity
equals the heat production rate. This metastable state has a long
life-time, of the order of the star life-time, and it ends when
the nuclear fuel is burned out. In addition, the quasi-equilibrium
is only local, since the temperature decreases from core to
surface. However, nuclear reactions are often, but not always,
sufficiently slow on the scale of thermal and mechanical exchanges
and take place on such a small scale that spatial and temporal
deviations from equilibrium can be neglected to a very good first
approximation.

At least in one limit the MB distribution can be rigorously
derived: systems that are dilute in the appropriate variables,
whose residual interaction is small compared to the one-body
energies. In spite of the fact that the effects of the residual
interaction cannot be neglected (the electron screening factor is
a well-known example of correction due to the astrophysical plasma
environment) at zero order the many-body correlations can be
neglected and the stellar interior can be studied in this dilute
limit. In this limit the velocity distribution is the Maxwellian
one.

However, one should keep in mind that derivations of the
ubiquitous Maxwell-Boltzmann distribution are based on several
assumptions~\cite{Coraddu:1998yb}. In a kinetic approach, one
assumes (1) that the collision time be much smaller than the mean
time between collisions, (2) that the interaction be sufficiently
local, (3) that the velocities of two particles at the same point
are not correlated (Boltzmann's Stosszahlansatz), and (4) that
energy is locally conserved when using only the degrees of freedom
of the colliding particles (no significant amount of energy is
transferred to collective variables and fields). In the
equilibrium statistical mechanics approach, one uses the
assumption that the velocity probabilities of different particles
are independent, corresponding to (3), and that the total energy
of the system could be expressed as the sum of a term quadratic in
the momentum of the particle and independent of the other
variables, and a term independent of momentum, but if (1) and (2)
are not valid the resulting effective two-body interaction is
non-local and depends on the momentum and energy of the particles.
Finally, even when the one-particle energy distribution is
Maxwellian, additional assumptions about correlations between
particles are necessary to deduce that the relative-velocity
distribution, which is the relevant quantity for rate
calculations, is also Maxwellian.

In the following we give arguments and mechanisms  that lead to
distribution functions that are different from the MB one in a
stationary state.

Correlations between particles, so that the probability
distribution of the system is not described by the product of
independent probabilities of the components, are in general
responsible for such more general distributions. The specific
microscopic mechanisms that generate these correlations depend on
the particular system and there exist many approaches to derive
the relevant distributions.

In an approach that uses the Fokker-Planck equation, which takes
into account the average effects of the environment through the
drift $J(p)$ and diffusion $D(p)$ coefficients, stationary
solutions different from the Maxwell distribution (e.g.
Druyvenstein or Tsallis like distributions) can be obtained, when
$J(p)$  and  $D(p)$ include powers of $p$ higher than the lowest
order
\cite{KaniadakisQuarati:1993e1997}. The presence of higher powers
of $p$, i.e., higher derivative terms, can be interpreted as a
signal of non-locality in the Fokker-Planck equation. We stress
that these distributions are stationary (stable or metastable) and
what counts to decide the distribution is the type of collisions
between ions and the dependence on momentum of the elastic
collisional cross sections (Coulomb, screened Coulomb, enforced
Coulomb, among others), or the presence of ion-ion correlations
\cite{newferro}.

The presence of random fields (e.g., distributions of random
electric micro-fields or, in general, of random forces) introduces
in the kinetic equations factors whose effect is to enhance or to
deplete the high-momentum tail of the distribution function
\cite{Coraddu:1998yb}.

Because of the many-body nature of the effective forces, which
makes the collisions not independent, the distributions of the
relevant degrees of freedom observed, e.g., the ones selected by a
fusion reaction, can be different from the distributions of the
quasi-particles that describe the plasma. In addition the plasma
makes effective interactions time dependent (memory effects) and
non-local. These effects depend strongly on the energy of the
selected particles and on the collisional frequency.

One important and clear example of this last point is given by the
fact that many processes, such as nuclear fusion itself, depend on
momentum rather than on energy. This distinction is important
because, due to plasma many-body effects, an uncertainty relation
holds between momentum and energy \cite{Ga:67}. Even when the
energy distribution maintains its Maxwellian expression, the
momentum distribution can be different in the high energy tail. In
fact, this quantum uncertainty effect (not Heisenberg uncertainty)
between energy $\cal{E}$ and momentum $p$, caused by the many-body
collisions and described by the Kadanoff-Baym equation, implies an
energy-momentum distribution of the form
\begin{equation}
f_Q({\cal E},p)=\frac{1}{\pi} n({\cal E}) \delta_{\gamma}({\cal
E},p)
\end{equation}
with
\begin{equation}
\delta_{\gamma}({\cal E},p)=\frac{Im \Sigma^R({\cal E},p)}{({\cal
E}-{\cal E}_p- Re \Sigma^R({\cal E},p))^2+(Im\Sigma^R({\cal
E},p))^2}
\end{equation}
where $\Sigma^R({\cal E},p)$ is the mass operator of the
one-particle Green function. After integrating in ${d\cal E}$ the
product of $f_Q({\cal E},p)$ and the Maxwellian energy
distribution, we obtain a momentum distribution with an enhanced
high-momentum tail. Although this approach produces a deviation
from MB distribution, the state represented by $f_Q(p)$ is an
equilibrium state \cite{St:00,Coraddu:2004cp}. The Maxwellian
distribution is recovered in the limit when $\delta_{\gamma}({\cal
E},p)$ becomes a $\delta$ function with a sharp correspondence
between momentum and energy.

Distributions different from the Maxwellian one can also be
obtained axiomatically from non-standard, but mathematically
consistent, versions of statistical mechanics that use entropies
different from the Boltzmann-Gibbs
one~\cite{Tsallis:1987eu,Kaniadakis:2004rj}.

We have argued that it is not sufficient to know that the
Maxwellian distribution is a very good approximation to the
particle distribution. We must be sure that there are no
corrections to a very high accuracy, when studying reactions that
are highly sensitive to the tail of the distribution, such as
fusion reactions between charged ions. Several mechanisms have
been outlined (others need to be studied) that can produce small,
but important deviations in the tail of the distribution.

\newpage

\begin{thebibliography}{99}
\bibitem{Atkinson:1929}
R.~d.Escourt Atkinson and F.~G.~Houtermans,
 Z. Physik {\bf 54} (1929) 656.

\bibitem{Castellani:1996cm}
V.~Castellani et al,
  Phys.\ Rept.\  {\bf 281} (1997) 309
  [arXiv:astro-ph/9606180].

\bibitem{Clayton:1968}
D.~Clayton, {\em Principles of Stellar Evolution and
Nucleosynthesis} (New York: McGraw-Hill Book Company, 1968).


\bibitem{Tsallis:1987eu}
  C.~Tsallis,
  J.\ Statist.\ Phys.\  {\bf 52} (1988) 479.

\bibitem{Coraddu:2001ps}
M.~Coraddu et al,
  PhysicaA {\bf 305} (2002) 282
  [arXiv:physics/0112018].

\bibitem{Degl'Innocenti:1998dy}
S.~Degl'Innocenti et al,
  Phys.\ Lett.\ B {\bf 441} (1998) 291
  [arXiv:astro-ph/9807078].

\bibitem{Coraddu:1998yb}
M.~Coraddu et al,
Braz.\ J.\ Phys.\  {\bf 29} (1999) 153 [arXiv:nucl-th/9811081].









\bibitem{KaniadakisQuarati:1993e1997}
  G.~Kaniadakis and P.~Quarati,
  PhysicaA {\bf 192} (1993) 677;
%
{\em ibid.} {\bf 237} (1997) 229.

\bibitem{newferro}
F.~Ferro and P.~Quarati,
Phys.\ Rev.\ E {\bf 71} (2005) 026408 [arXiv:cond-mat/0407665].

\bibitem{Ga:67}
V.~M.~Galitskii and V.~V.~Yakimets,
JEPT {\bf 24} (1967) 637.
%
%
\bibitem{St:00}
A.~N.~Starostin, V.~I.~Savchenko, and N.~J.~Fisch,
Phys. Lett. A {\bf 274} (2000) 64.

\bibitem{Coraddu:2004cp}
M.~Coraddu et al,
  PhysicaA {\bf 340} (2004) 490
  [arXiv:nucl-th/0401043].

  \bibitem{Kaniadakis:2004rj}
G.~Kaniadakis et al,
  Phys.\ Rev.\ E {\bf 71} (2005) 046128
  [arXiv:cond-mat/0409683].

%
\end{thebibliography}
\end{document}